\documentclass[final,3p,times]{elsarticle}

\usepackage{graphicx,amssymb,amsmath,color,float,mathtools}
\usepackage{enumitem}
\usepackage{inputenc}
\usepackage{blindtext}
\usepackage{booktabs,multirow}
\usepackage{color,soul}
\usepackage{amsthm,mathrsfs,amsfonts,dsfont,bm}
\usepackage{tasks}
\usepackage[export]{adjustbox}

\usepackage[english]{babel}
\usepackage{hyperref}
\usepackage{subcaption} 
\journal{Physica A: Statistical Mechanics and its Applications}
\begin{document}
\begin{frontmatter}
	\title{Differences in structure and dynamics of networks retrieved from dark and public web forums}
	\author{Maryam Zamani,$^1$ Fereshteh Rabbani,$^1$$^,$$^2$ Attila Horics\'anyi,$^1$ Anna  Zafeiris,$^3$ Tamas Vicsek,$^1$$^,$$^3$\\
		{\small $^1$Department of Biological Physics ,  E\"otv\"os University, P\'azm\'any P. stny 1/A, 1117, Budapest, Hungary
		} \\
		{\small $^2$Department of Physics, Shahid Beheshti University,
			G.C., Evin, Tehran 19839, Iran} \\
		{\small $^3$MTA-ELTE Statistical and Biological Physics Research Group, Pázmány Péter s. 1/A, 1117 Budapest, Hungary}\\}


\begin{abstract}
Humans make decisions based on the information they obtain from several major sources, among which the comments of others in Internet forums play an increasing role. Such forums cover a wide spectrum of topics and represent an essential tool in choosing the best products, manipulating views or optimizing our decisions regarding a number of aspects of our everyday life. However, many forums have extremely controversial topics and contents including those which radicalize the readers or spread information about dangerous products and ideas (e.g., drugs, weapons or aggressive ideologies). These just mentioned activities are taking place mainly on the so called “dark web” allowing the hiding of the identity of members using dark forums. We use network theoretical approaches to analyze the data we obtained by studying the connectivity features of the members and the threads within a wide selection of forums (including dark and semi-dark) and establish several characteristic behavioral patterns. Our findings reveal both common and rather different features in the two types of behavior. In particular, we show that the various distributions of quantities, like the activity of the commenters, the dynamics of the threads (defined using their lifetime) or the degree distributions corresponding to the three major types of forums we have investigated display characteristic deviations. This knowledge can be useful, for example, in identifying an activity typical for the dark web when it appears in the public web (since the public web can be accessed and used much more easily).
\end{abstract}


\end{frontmatter}
\section{Introduction}
Internet has a huge amount of data regarding many kinds of items. Data generated by people reflecting their interests or opinions are of particular importance from various viewpoints \cite{Zafarani,Contijoch}. In fact, this is one kind of the manifestations of the so called "wisdom of crowds" effect \cite{Surowiecki}. Among the most relevant applications of such corpora are the various recommendation systems  (see, e.g., \cite{Montaner,ChapterBook}) some of which provide direct suggestions for buyers of an Internet store, while some others provide very different services. Consider for example the case of Cambridge Analytica \cite{cambridge_Analytica} which sold data - downloaded from Facebook - for a large sum, and these data were later used for political purposes. There are some further important possible applications of data "scraped" from Internet forums: they both give an insight into the way people prefer to interact regarding a given topic, but it also gives a possibility - in principle - to track down radicalizing content, or even radical/terrorist groups \cite{Skillicorn,Reid}.
Due to recent developments nearly all social media sites have been able to prevent access to their personal pages (except, of course, for those who got individual permission from the owner of the page). This makes the study of forums a reasonable choice, since most of them are open to the public.
Forums represent an important part of the web and the increase in their popularity and importance have made them a subject of recent studies \cite{Himel,Hong, Holtz}. Forums are typically moderated. They consist of so called "threads" which are made of the comments by the users, related to a single "first entry" or “root comment” about the given subject. Users/commenters, as a rule, use nicknames or IDs in order to identify themselves. These IDs are not sufficient for tracking back a person’s name, but within a public forum, in principle, the IP number of a user can be recovered. We shall discuss this aspect later. The huge increase in the number of users of forums as time goes on have lead researchers to study the type and structure of the resulting interactions \cite{Musial}. Many statistical models have been developed \cite{Aragon, Sprent, Nishi} for characterizing the dynamics of online discussions \cite{WangC,Backstrom, Gomez} and understanding the structure of threads and the resulting networks \cite{GomezV}. Some of them have been aiming at the prediction of the dynamics of threads and user behaviour \cite{Medvedev, Backstrom}. Techniques are proposed for reconstructing the structure of forums \cite{Aumayr} by finding the pairs of answers and questions \cite{Cong,Ding,Yang} and identifying the focus of a conversation in threads using National Language Processing (NLP) methods \cite{Feng,Chan}. By using sentiment analysis techniques, the polarity of posts is determined and opinions of users with multiple languages have been classified \cite{Park,Abbasi}. Woo and Chen studied the diffusion of information in forums using an epidemiology model \cite{Jiyoung}. 
In this paper we shall consider the surprisingly large spectrum of Internet forums, provide their classification, analyze their most important features and compare their characteristics. Social networks could be easily extracted from all types of websites in which users communicate and interact with each other online, if the data regarding their communication were available. However, the access to such personal data is highly restricted by the social media website providers. One way of getting around this limitation, as mentioned above, is studying forums. The way of extracting networks can be based on either direct communication or indirect interactions, such as leaving comments on the same posts or threads. These types of networks can be formed on the dark web as well \cite{Everton} where, for different reasons, people want to stay anonymous and want to make the tracking of their activity and identity impossible. Many publications refer to forums as being dark because of the radical nature of their content, even if they are available from public search engines such as Google \cite{Zhang}. In this paper the usage of the term dark web refers only to the domain of .onion web. Griffth et al. \cite{Griffith} analyzed the hyper-link topology of the dark web (.onion domain) and found extreme dissimilarity compared to the public web. For example, they have found that more than 87\% of dark web sites do not connect to any other site. Domenico et al. \cite{Domenico} studied the structure of the dark net as well, and they found it to be more resilient compared to the public net, with respect to random failure. Yet, there is still a lack of knowledge regarding the characteristic features of the structure of interactions among commenters and threads inside dark web forums.\\
Thus, our paper is the first one dedicated to studying the hierarchical network structure and further properties of dark web forums, extracted from the .onion web domain and comparing the resulting user interaction networks with the ones formed in public webs \cite{Barabasi}.
Our main goal is to reveal how the unidentifiable nature of the commenters affect the structure of the emerging networks and, in general, the online behavior of interactions among dark web users (as compared to those using the public web).
\section{Forums} 
In this section, we give a brief description of the types of forums we have used and classify them based on their accessibility, content and structure. We scraped forum-data from both public, dark and semi-dark (see below) web sites using the Scrapy Python library \cite{Scrapy}.
\begin{itemize}
\item Accessibility: the forums we have analyzed fall into the following three categories based on their accessibility:
\begin{enumerate}
\item \textit{Public forums}, which are accessible through the usual web search engines, such as, e.g., Google and Bing. A widely popular example for a public forum is Reddit which is diverse in terms of topics and rich in terms of both quantity and quality and has been a popular forum for studying its features see, e.g., references \cite{MedvedevA, Gilbert, Glenski, Newell}. Further examples include Naira-land which is a forum for discussion on a variety of topics from food to politics, and The Wholesale Forums (TWF) which is a marketplace for wholesalers, importers, drop shippers, retailers and trade buyers.
\item \textit{The dark web} \cite{Jamie}, referring to the part of Internet in which the hosted websites are accessible only through some special software that disguises IP addresses. The most common software used to access the dark web is The Onion Router, referred to as TOR. This part of web is not indexed by normal search engines and usually used for illegal purposes like buying credit card numbers, guns, drugs etc \cite{Cath}. We have scraped forums from different dark websites such as, Dream Market, which has been operating in illegal marketing; Pedo Support Community which is related to Pedophilia and child abuse; and AnonGTS which is a forum dedicated to the "art" of pornographic photos and cartoons. 
\item \textit{Semi-dark web forums} is an expression which we use to refer to forums initiated in the public domain but later on, for some reason, are moved to the dark web (or back and forth). An example of this type of forum is 8chan which is an image board style forum that does not restrict the content of the posts. After a while, due to its contents related to child pornography, swatting activity and the Gamergate controversy, it has been removed from Google search engine and moved to the dark web. In 8chan, there are different types of categories with topics from politics to pornography. In the ones with seedy contents, the users’ IP addresses  are anonymous like in the dark web, that is why we refer to 8chan forum as semi-dark in our paper.
\end{enumerate}
\item Structure: Forums are structured in different ways. We used both the IDs associated with users (who as a rule choose such IDs for themselves) or the IDs of threads mostly given by the forum service provider. Some of forums have a hierarchical structure in which each comment is a direct reply to a comment posted beforehand (which can be the root comment as well). In such cases the resulting networks are directed with different number of hierarchical levels. In some other types of forums, it is not clear which user responds to whom, in these cases we used the thread ID to construct the corresponding network (see section \ref{sec:NetCr}).  
\end{itemize}

\section{Creating Networks}
\label{sec:NetCr}
We have created two types of interaction networks (directed and undirected) from the forum data as follows:\\
\subsection{Comment and commenter graphs}
This is for the case when in the forum data, it is clear who replied whom. Two types of networks could be extracted: (i) In the first one the nodes are the users (with their corresponding IDs) who are connected with a directed edge if one (target) replies to the post of the other (source). The resulting graph is directed and weighted, the weight reflects the frequency of the replies. (ii) In the second case, the nodes are the comments which are connected if one is a reaction (answer) to the other. In this case the former one is the "child" comment and the latter is the "parent" comment. Assuming that each post has (maximum) one parent comment, the resulting graph is a tree.

\subsection{Co-commenting Network} 

\begin{figure}
\centering
\includegraphics[width=0.9 \textwidth]{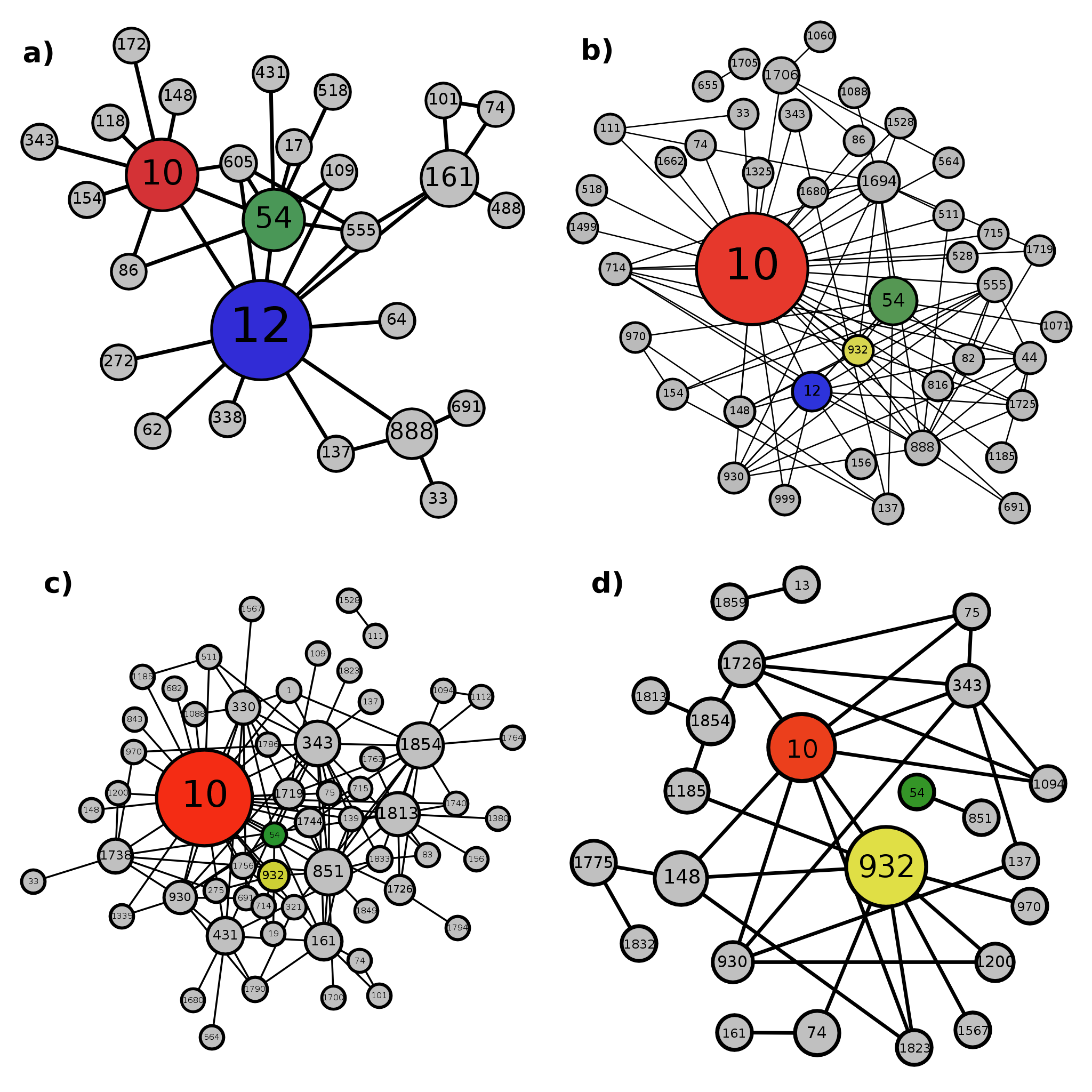}
\caption{\label{fig:222} Network’s dynamics visualized for four different time intervals in the order of increasing date (see text). The radius of the nodes is proportional to the values of betweenness centrality. Nodes with higher betweenness centrality are bigger.
}
\end{figure}

For the case when it is inaccessible who replied whom in the forums, we made networks based on bipartite graphs. Two users are connected if they make comments on the same threads. The resulting network is undirected and weighted. The strength of interaction is higher if two users share more threads. Users who are mostly active and make comments in many different threads inside a forum are important in terms of propagating their own idea or ideas of others. We can have networks of threads as well. Two threads are connected if at least one user makes a comment on both of them. The resulting undirected networks are weighted and weights are defined based on the number of users (linearly proportional to this number) who made comments on both threads. In the present paper, we considered  a limited set of bipartite networks leading to the results shown in Figure \ref{fig:222}.  
\begin{figure}
\centering
\includegraphics[width=0.8\textwidth]{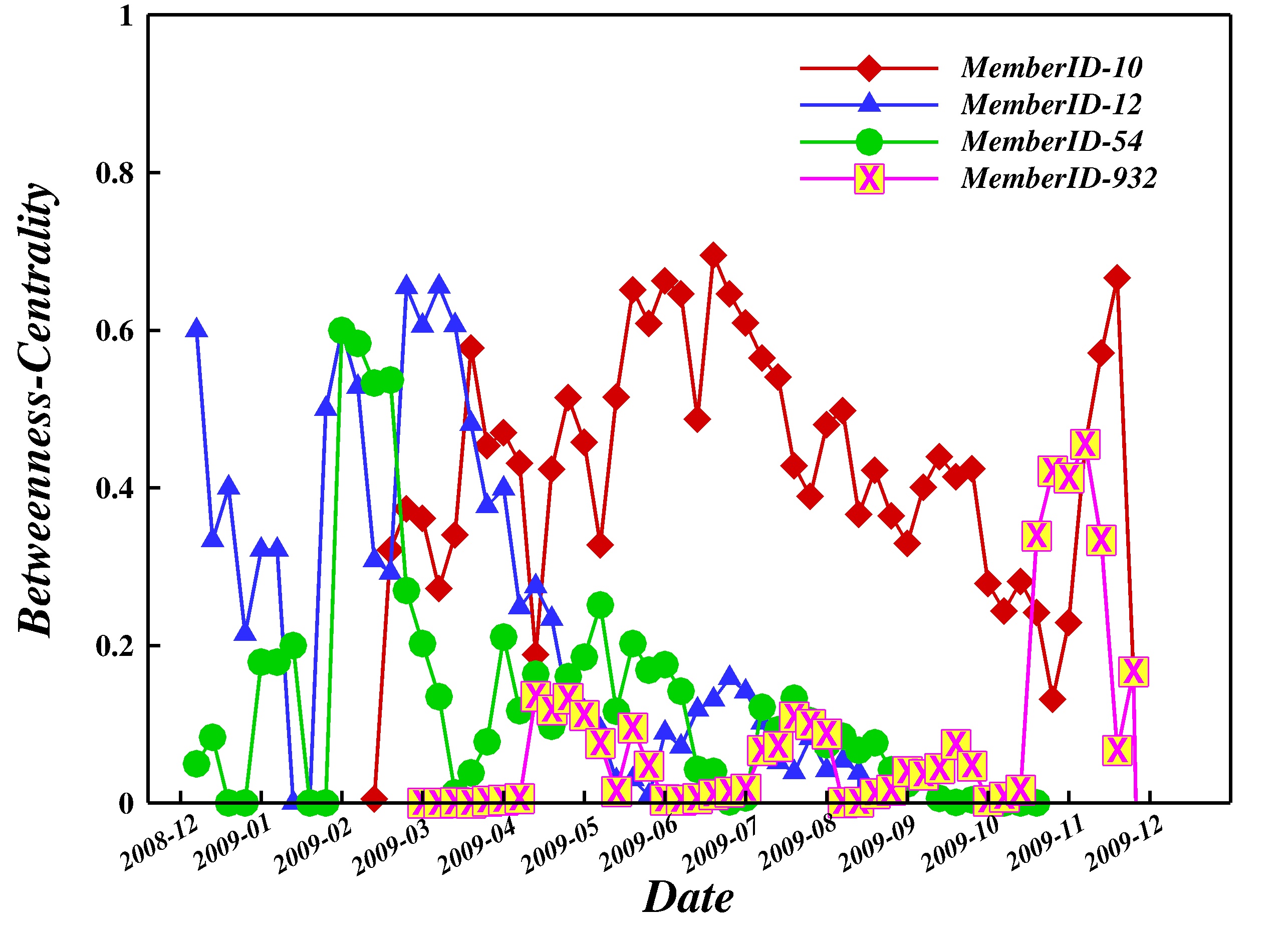}
\caption{\label{fig:BC-Ansar_Bipartite} The change in betweenness centrality over time for the four most active users depicted in Figure \ref{fig:222}.}
\end{figure}
\subsection{Evolving Network} 
Real systems are usually not static, instead they evolve in time \cite{Barzel}.
This can be manifested in the emergence of new parts, disappearance or merging of existing parts, and the relations among constituents can be rearranged over time as well. Therefore, the network representation we apply should be able to model the time evolution of the underlying system. The dynamics of networks extracted from Ansar forums \cite{Ansarforum} in four different time steps are shown in Figure \ref{fig:222}. In these graphs two nodes (users) are connected if they make comments at least on two common threads over a period of one month. If a node (user) has not been active for one month, it is simply removed from the graph. The radius of the nodes in the graphs is proportional to the value of their betweenness centrality. The corresponding dynamics of the value of betweenness centrality for four most active nodes in the graphs of Figure \ref{fig:222} are depicted in Figure \ref{fig:BC-Ansar_Bipartite}. Figure \ref{fig:222}-a demonstrates the network structure in the period between 2009-2 and 2009-3 of Figure \ref{fig:BC-Ansar_Bipartite}, in which the node with member-ID 12 has the largest value of betweenness centrality. The structure of the network changes in time, new nodes (corresponding to new individuals who are participating in conversation) are added to the graph. Node number 10 becomes more active by making comments in more threads so it gets higher betweenness centrality.
On the other hand, node 12 lost its importance as displayed in Figure \ref{fig:222}-b, its betweenness centrality decreases and finally in Figure \ref{fig:222}-c completely disappears from the graph. Figure \ref{fig:222}-c corresponds to the period between 2009-9 and 2009-10, when node 54 gradually loses its importance and at last in Figure \ref{fig:222}-d it is separated from the big component of the graph. At the same time node 932 acquires higher betweenness centrality by becoming more active and making comments in more threads.

\section{Results}
\subsection{Comparing comment graphs in the public, semi-dark and dark webs}

Ideally, a comparison would involve forums having the same or similar topics. Radicalizing forums would be of particular importance from the point of global safety. However, such forums are hardly accessible. Even if one is able to locate such a forum on the dark web, even the reading of it needs membership and the membership requires invitation by someone, who is already a member. Obviously, standard (public, done by regular scientists) research cannot meet this condition. Furthermore, there are additional methods of hiding. For example, the URL-s corresponding to given dark web pages may change frequently, e.g., following an algorithm which is known only by a very small circle (so that even a dark web search engine is unable to locate the terrorism-related URL. Due to these barriers the range of dark web forums we can analyze is limited. In fact, some of the prior studies about dark web forums eventually involved investigating terrorism-related forums accessible through the public web.

\begin{figure}
\centering
\begin{subfigure}{.9\textwidth}
  \centering
  \includegraphics[width=0.95\textwidth]{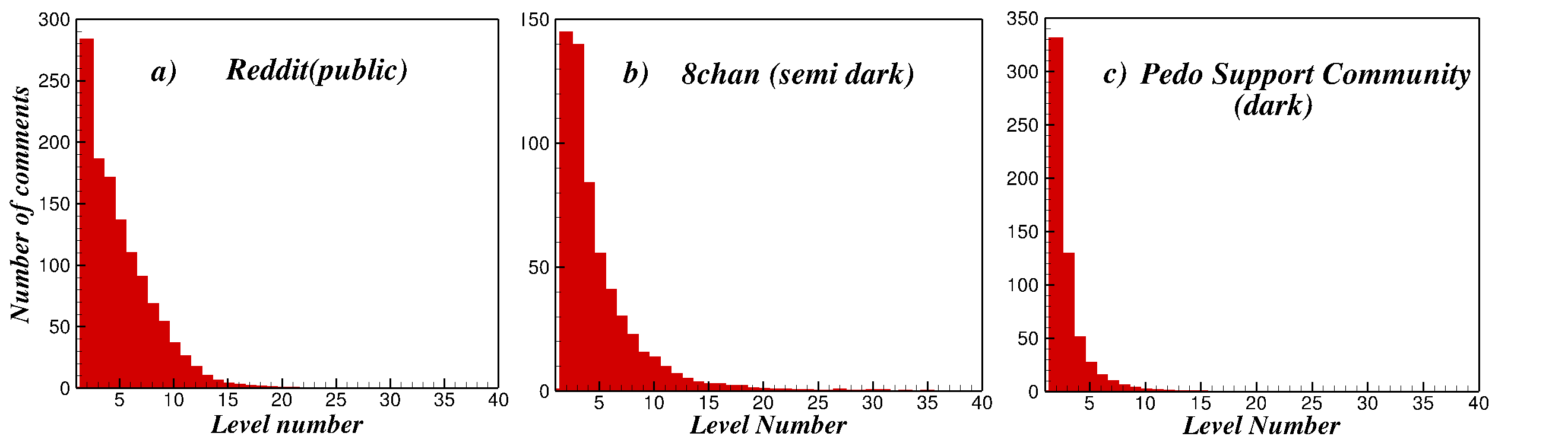}
  \end{subfigure}
  
\begin{subfigure}{.9\textwidth}
  \centering
    \includegraphics[width=0.95\textwidth]{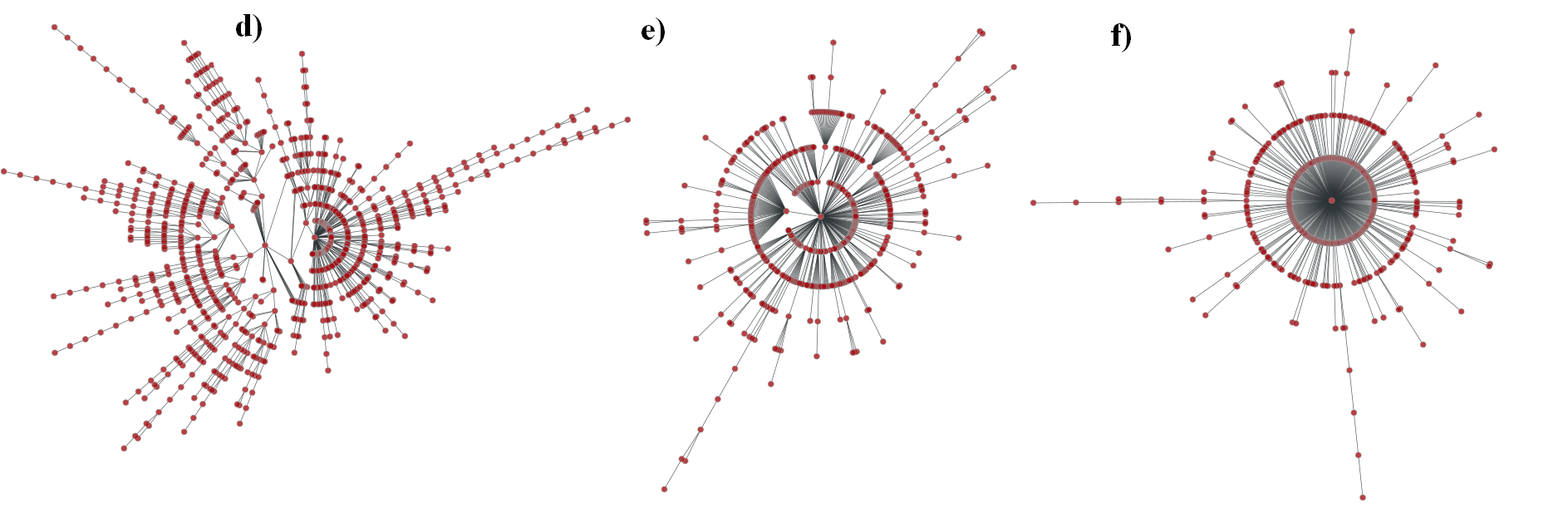}
  \end{subfigure}
\caption{\label{fig:Conv-depth}
Histograms demonstrating the "depth of the conversation" for the three web domains (a) public, (b) semi-dark and (c) dark, respectively. Conversations on the public and semi-dark domains often contain longer chains of posts responding to each other resulting in deeper conversations, while discussions in the dark web tend to be more superficial. Comment graphs are trees by definition, thus, they can be best visualized by using programs having an option for layered output. Networks (d), (e) and (f) are typical representations of the comment graphs corresponding to the domains for which the distributions are shown above them. The histograms depict the averages of ten threads downloaded from the Reddit, 8chan and Pedo Support Community forums, respectively.}
\end{figure}

One part of our data comes from Dream Market and Pedo support community (as for dark web), another part comes from political forums downloaded from 8chan website (as for the semi-dark), and finally we have Reddit (as for the public). In order to have the most reasonable comparison, where available, we have looked for similar topics (politics and political related threads from public and semi-dark web) with similar size (containing minimum 500 posts). In the case we have not found reasonably large forums in the same topics, the analysis was done for more or less the same size forums in related topics or sometimes completely different topics. 
By studying various features related to the (i) threads, (ii) the comment graphs and (iii) the behavior of users, we have found differences in the characteristics of the threads and in user behavior. Threads have many basic characteristics, such as number of posts (which defines their length), number of commenters, lifetime, etc. Among these, the most striking difference is related to the length: while threads in publicly available forums can easily exceed one thousand posts or more, such lengths are rare in dark web forums which usually comprise a few tens or a few hundreds of comments.
Another important feature characterizing a thread is the "depth of the conversation". We have analyzed this property in the following way \cite{RedditAnalBlog}: a number can be assigned to each comment determining its ’level’, showing its distance from the root comment. Accordingly, the level-number of the root comment is 0, the level-number of the immediate posts reacting on the root comment is 1, and, in general, the level-number of a post reacting on a comment on level $n$ is $n+1$. In the theoretical case in which each comment is a reaction to the preceding post, each level would contain exactly one comment, that is, the distribution of the level-numbers would be constant equal to 1. This case describes a deep conversation. In the other extreme theoretical case, in which each comment is a reaction to the root comment, the 0th level would contain exactly 1 post (the root comment) and all other posts would have 1 as their level-number. This would indicate a very superficial (depthless) conversation. Real-life threads are somewhere in between.  Figure \ref{fig:Conv-depth} depicts histograms for the three domains (public, semi-dark and dark). Sub-figures \ref{fig:Conv-depth}-a and \ref{fig:Conv-depth}-b are related to political topics from the public Reddit and from the semi-dark 8chan, respectively. As it can be seen, these threads contain longer chains of comments responding to each other. The Pedo Support Community (depicted on sub-figure \ref{fig:Conv-depth}-c) is devoted to sharing experiences related to a certain topic, such as first pedophile affairs, jail-time or favorite dresses on children. Although these topics could result in deep conversations, our results indicate that people who are active on this forum are more motivated in writing memoirs than conferring about the given topic. 
And this leads us to our next observation, namely to the differences in \textit{user activity}. This property is defined by the number of comments posted by a certain user. People are most active in the dark web (where on average one user posts $7.8$ times on a thread), then in the semi-dark web, where this number is $3.79$ (that is, half as much) and finally, people leave the fewest comments in public forums, where this average is only $1.78$. These differences are partly due to the fact that the most active users are much more active in the semi-dark and dark domains than in the public web: while the number of comments posted by the most active user in the public forum is around $15$, this number in 8chan is larger than $100$ and in the dark forums it often exceeds even $200$. This phenomenon, in the dark web, is due to the high activity of a few users whose aim is to sell or advertise some illegal products, and in 8chan, it might reflect some kind of political fanaticism. Another property by which users can be characterized is their influence. We define influence based on the number of immediate comments a user evokes. Of course, since more posts can trigger more reactions, more active users have a better chance to become more influential. For example, if the posts of a certain user evoke $20$ comments out of $1000$ (which is the full length of the tread in this example) then the influence of this user will be 2\%. Considering a user as being "influential" in case she/he evokes minimum 1\% of all the comments within the thread, we find that the number of influential users is more than twice higher on the semi-dark web than on public forums (averages are $18$ vs. $7.9$). This observation is in contrast with our expectations since it implies a more distributed opinion leadership on 8chan forums than on public ones. The value measured on dark web forums is the smallest ($6.7$ influential users on average in each thread), and as such it is in line with our expectations since it reflects homogeneously distributed opinion leadership.
\subsection{Behavioral dissimilarities in different web domains}
\begin{figure}
\centering
\includegraphics[width=0.60\textwidth]{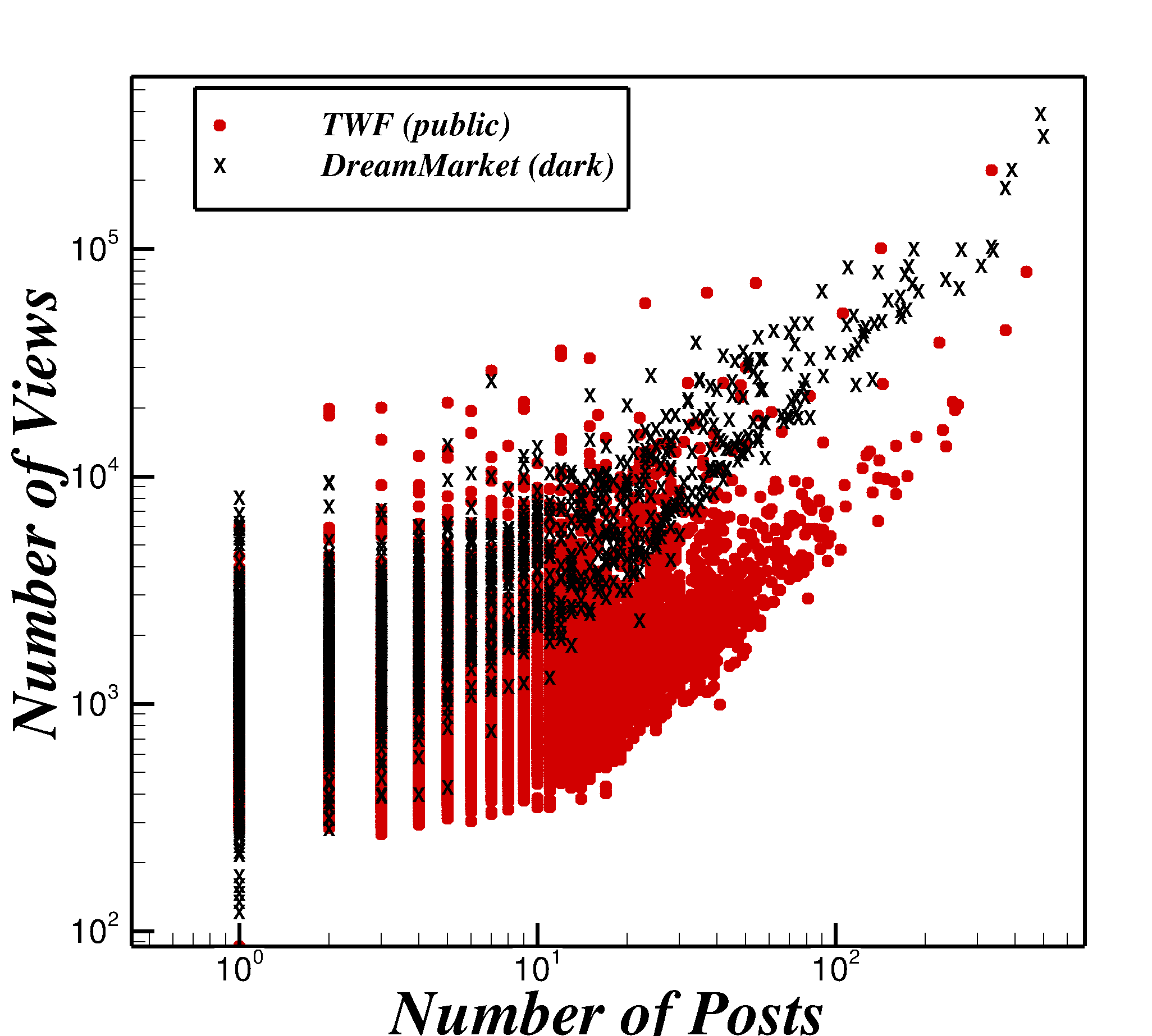}
\caption{\label{fig:post_vs_views} Log-log plot of the number of views as a function of the number of posts. Each dot in the figure represents a thread: the x axis shows the length of the thread (number of posts), and the y axis refers to the number of people visiting the thread. Users of the dark web seem to prefer to leave without a trace, expressed by the fact that the view/post ratio is magnitudes higher on the dark web (note the scale on the y axis). Visitors of public forums make comments on a post more easily. Such statistical characteristic differences can be used to identify illegal or suspicious contents even without understanding the language.}
\end{figure}

Typical differences in statistical characteristics of forums can be used to categorize these websites automatically, even without understanding the language they are written in. Such a difference originates from the observation that people seem to leave comments more easily on the public than on the dark web, or, which is equal, many people seem to hesitate to leave a mark on the dark web. (Maybe because they feel it as an irrevocable illegal act?) Accordingly, the ratio between the number of visits (referring to the event when somebody clicks on the title of the thread) and posts (referring to the act of actually leaving a comment) differs considerably in the two kinds of web: while a visitor in a public forum leaves a message easily, visitors of dark web forums prefer to leave without any trace. Figure \ref{fig:post_vs_views} shows how users behave on a dark web forum called "Dream Market" (marked with black ’x’) and on a public forum named TWF (marked with red circles).
\begin{figure}
\centering
\includegraphics[width=0.95\textwidth]{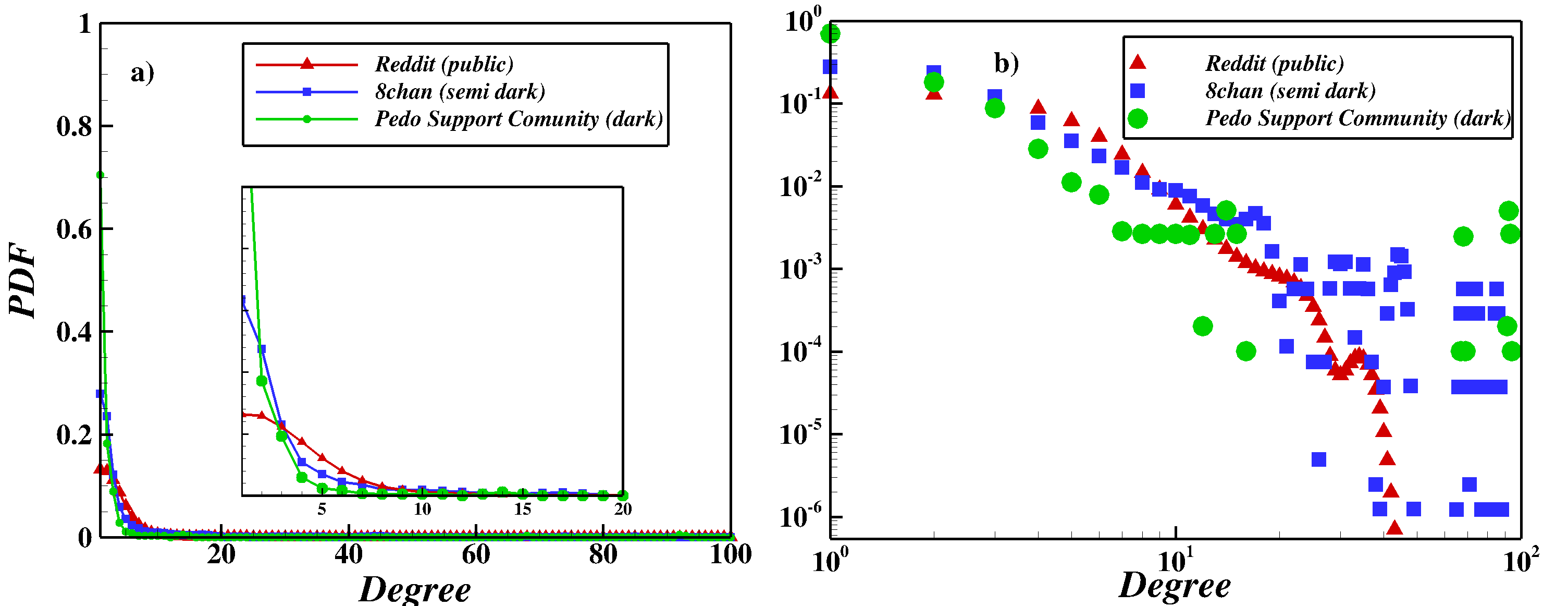}
\caption{\label{fig:DegreeDist} a) Degree distribution of networks constructed from forums in dark, semi-dark and public webs. b) Degree distributions in log-log scale indicate an exponential cutoff for the public and a slow decay for networks from dark and semi-dark forums.}
\end{figure}
\subsection{Network’s degree distribution in public, semi-dark and dark webs}
Figure \ref{fig:DegreeDist} demonstrates the degree distribution of commenter networks of forums from dark, semi-dark and public webs. The result has been obtained by averaging over 20 networks from different threads with almost the same number of nodes. The average number of nodes and edges in the networks of public threads are $N=127.7$, $M=267.3$ and for dark threads these are $N=85.33$, $M=117.66$. The degree distribution in public forums is more homogeneous than in semi-dark and dark ones. The number of users in public forums is much higher and they are almost equally more active. Therefore, the resulting network is bigger with higher number of nodes and edges. Since users make comments on public forums actively and more or less equally, it leads to a homogeneous degree distribution. In contrast, there is a limited number of users in dark forums where just few of them (for instance, sellers in dark web markets) are more active and have the highest degrees. 80\% of nodes in networks of dark forums have degree one. The difference between degree distribution of networks from these three web domains are more obvious in their tails. There is a constant decrease in the degree distribution of networks from public threads, while degree distribution in the networks from semi-dark and dark threads show a slower decrease, especially for the dark ones, which is due to the high activity of few users. Despite user’s behavior in public, semi-dark and dark forums, the resulting difference is also due to the structure of forums. In dark web forums each user can make comments in more than one post at the same time, or in other words each comment could have more than one parent comment. The average degree in networks from public threads is around $3.5\pm 0.1$, and in dark web is $3.0 \pm 0.1$. However, most users in dark networks have degree one or two and these are the highly active users who pull up the average degree by making many comments. According to the clustering coefficient (normalized to the same size of Erd\"os-R\'enyi graph) in networks of threads from public and dark forums, which are 1.3 and 0.60 respectively, dark networks are sparser with just a few nodes with high degree.

\begin{figure}
\centering
\includegraphics[width=1 \textwidth]{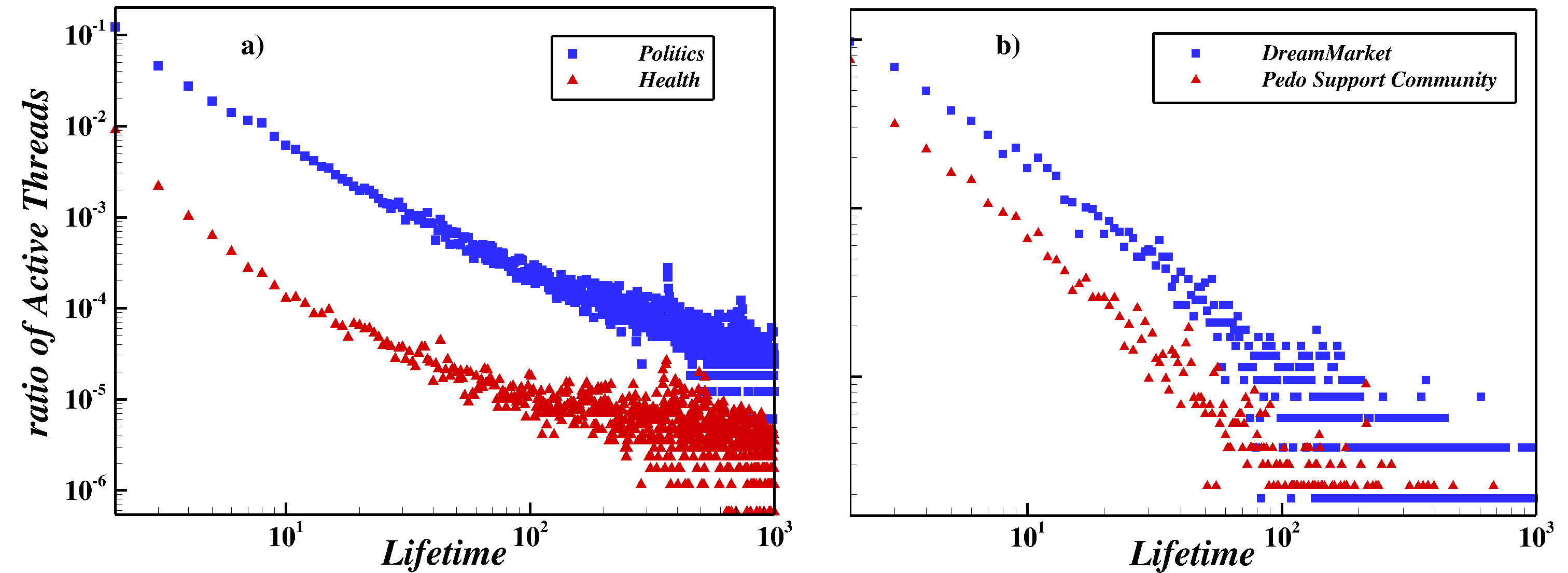}
\caption{\label{fig:Timelife} Ratio of active threads (the number of active threads at the corresponding lifetime divided by the whole number of threads in that forum) versus lifetime from two different topics (categories) in each public web (a), where the slopes of curves are smoother than in dark web, demonstrate the willingness of most users in public forums to be active and make comments. b) Dark web, curves show a fast decay as lifetime increases. Except for few overactive users in dark forums, others have less tendency to leave comments. Threads in dark forums live shorter.}
\end{figure}

\begin{figure}
\centering
\includegraphics[width=1 \textwidth]{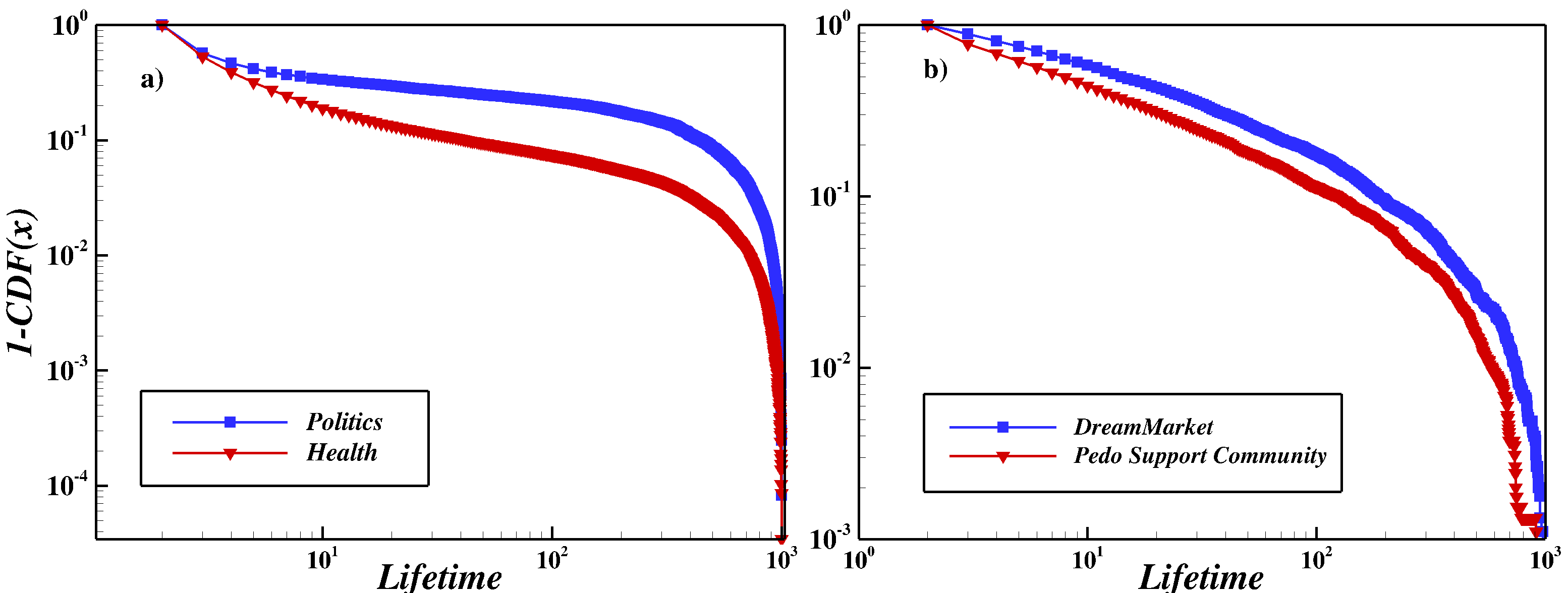}
\caption{\label{fig:Timelife2} $(1-CDF(x))$ versus life time, where $CDF$ is the cumulative distribution function of the values in Figure \ref{fig:Timelife} and x is the lifetime. a) Public forums. b) Dark forums. The difference between the distribution of active threads with various lifetime is more apparent in this figure.}
\end{figure}

\subsection{Thread’s lifetime in public and dark webs}

The other interesting question in comparing threads from dark and public forums is the difference between their lifetime, which is the number of days elapsed between the first and the last post in each thread. The ratio of active threads (the number of threads with corresponding lifetime has been divided with the total number of threads in that forum). This has been done for forums of two different categories (both in dark and public) versus their lifetime. The results are shown in Figure \ref{fig:Timelife}. 
As expected, there is a negative slope in the plots indicating that most threads have short lifetime (a few days), then, as lifetime increases, the ratio of active threads decreases. The size of the forums and the number of posts in public web are much larger than in the dark. The slopes of curves in Figure \ref{fig:Timelife}-a for public forums are smoother than those in Figure \ref{fig:Timelife}-b for dark. Due to the high number of users in the public forums and more importantly the tendency of most of them to make more than one comment, the ratio of active threads and their life-time decays slower than in dark forums. Although the user activity in dark forums is higher due to the high activity of just few users (who are over active), others have less tendency to remain active and leave more comments. Therefore, in dark forums there is a fast decay in the ratio of active threads as their lifetime is increased. These two kinds of behavior in public and dark are more apparent in Figure \ref{fig:Timelife2} which shows $1-CDF(x)$ versus x, where $CDF$ is the cumulative distribution function of the values in Figure \ref{fig:Timelife} and $x$ is the lifetime.

\subsection{Hierarchical structure of networks of forums}
\begin{figure}
\centering
\includegraphics[width=1 \textwidth]{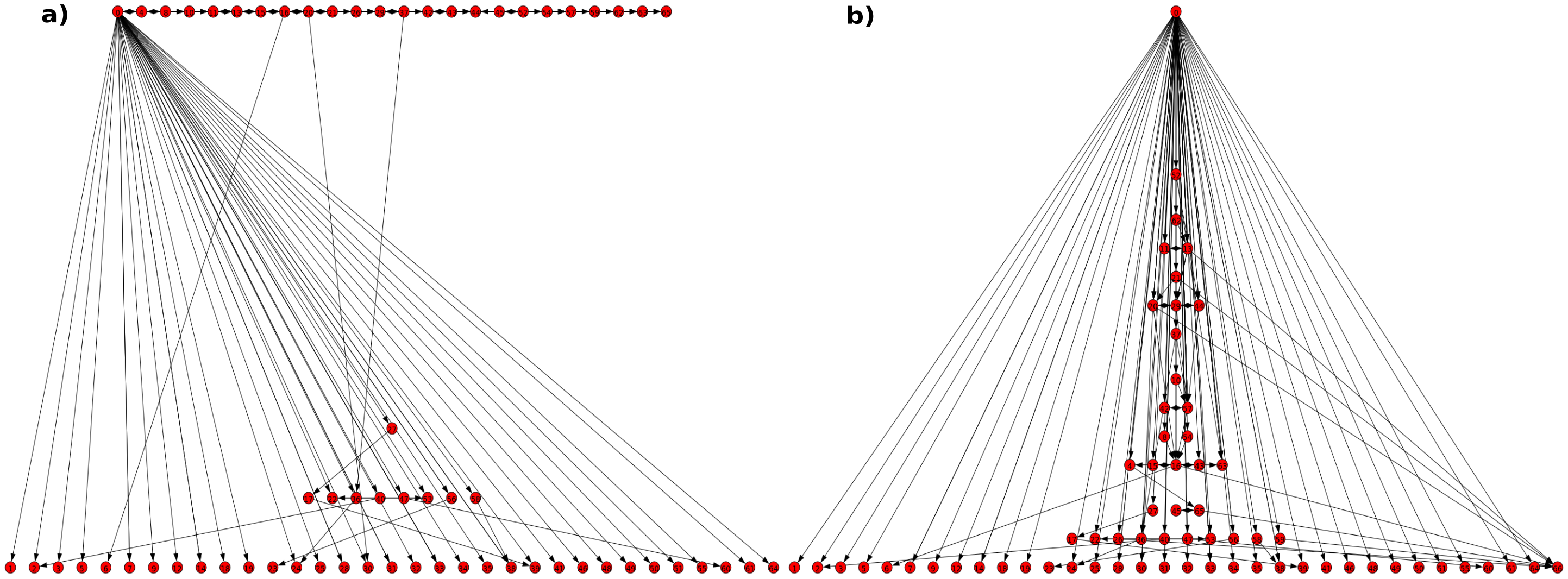}
\caption{\label{fig:Hierarchy} Hierarchical visualization of directed networks of commenters from a dark web forum (Pedo Support Community) with N=$66$ nodes and $110$ edges. a) Root commenter is active, participates in the thread and comments back to various posts from other individuals. The resulting network is less hierarchical with only three levels of hierarchy, the average clustering coefficient is $0.238$, average Path length=$2.8$ and GRC=$0.64$. b) Root commenter does not comment back on the other posts, the average clustering coefficient is $0.168$, average Path length=$1.88$ and GRC=$0.96$. Depending on the behavior of the root commenter, the structure of the resulting network could change significantly.}
\end{figure}

\begin{figure}
\centering
\includegraphics[width=1 \textwidth]{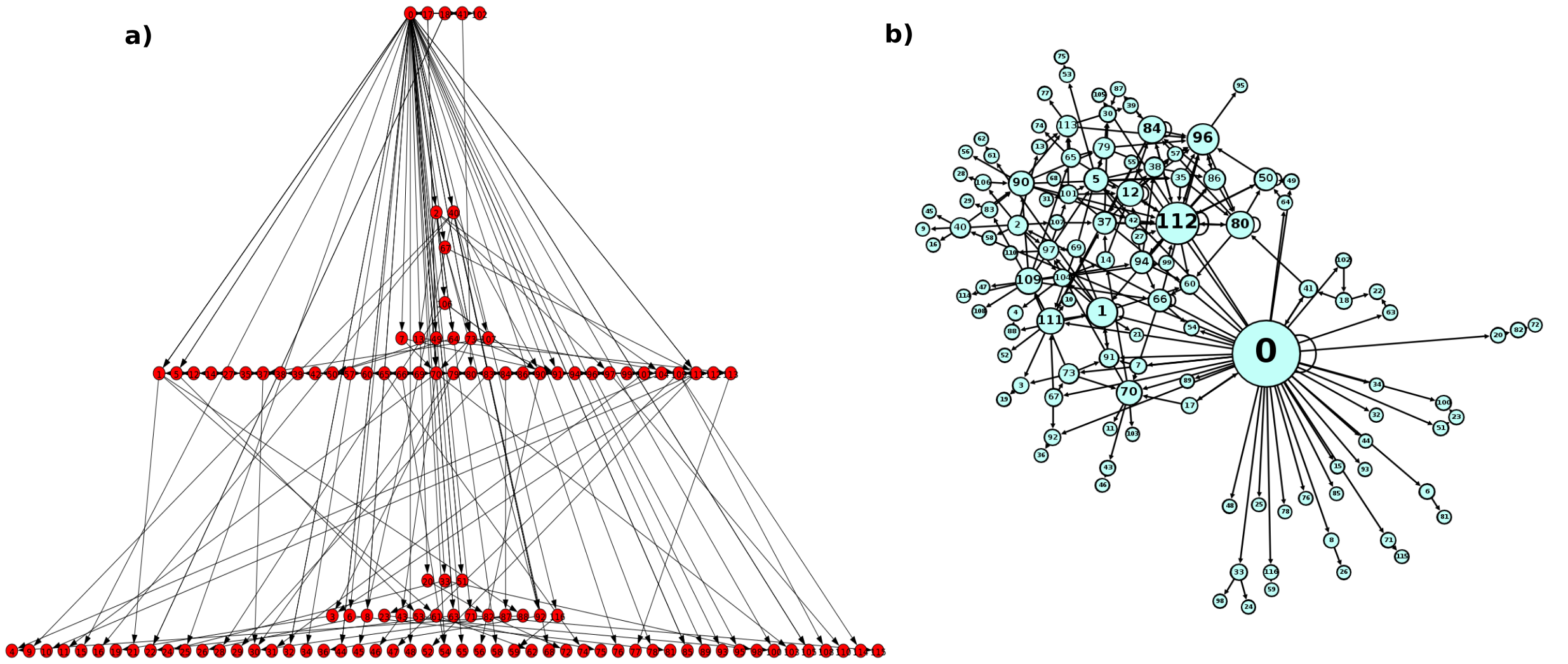}
\caption{\label{fig:Hierarchy3} Directed Network of commenters in semi-dark web (8Chan forum) with N=$117$ nodes and $214$ edges. a) Hierarchical visualization, root commenter comments on few other posts in the thread, average clustering coefficient is $0.047$ and GRC=$0.76$. b) Gephi visualization of panel (a)}
\end{figure}

Hierarchy is one of the most important features in nature \cite{Anna}, society and organizations \cite{Zamani}. In this section the hierarchical structure of commenter networks of forums from different web domains is studied and compared. The level of hierarchy can be measured by several methods including  Global Reaching Centrality (GRC) \cite{Mones}. GRC is calculated using the distribution of the local reaching centralities $C_R(i)$ which is proportional to the number of nodes that can be reached from node $i$ through the directed edges. Nodes on the top levels of hierarchy have higher local reaching centrality. GRC scores are between 0 and 1 and could be measured using the following expression,
\begin{equation}
GRC = \frac{\sum_{i\in V}[C_R^{max}-C_R(i)]}{N-1}
\end{equation}
where $C_R^{max}$ is the largest of $C_R(i)$, the summation is over all the nodes in the graph V and $N$ is the total number of nodes. 
In Figure.\ref{fig:Hierarchy}, the hierarchical structure of a dark web forum (Pedo Support Community) is displayed. Depending on the behavior of the root commenter, two types of structure could be observed. Root commenter is basically an individual who starts the conversation in a thread while the other individuals make comments on his/her post or on the comments of others. In case the root commenter is more active and starts to comment back or responds to others’ posts, the structure of the hierarchical networks will change. When root commenter responds to the high number of other individuals, the level of hierarchy decreases (Figure \ref{fig:Hierarchy}-a), in such a case there is a bi-directed link between the node who has been commented by the root commenter and the root commenter itself. As root commenter has higher local reaching centrality by accessing all nodes in the graph, the nodes who are the source of the directed links to the root commenter have also higher values of local reaching centrality and will be placed on the top of the graph. In case root commenter does not comment back, the structure of the graph is more hierarchical with GRC values exceeding $0.8$ (Figure \ref{fig:Hierarchy}-b). These types of structures are independent of the topics and web domains. In this section, we just present the ones from semi-dark and dark webs. Figure \ref{fig:Hierarchy3}-a is the hierarchical visualization for the network of semi-dark thread, where the root commenter comments back to just few other individuals. Figure \ref{fig:Hierarchy3}-b demonstrates the same network in the other visualization platform (Gephi) \cite{ICWSM09154}. In the caption of Figures \ref{fig:Hierarchy} and \ref{fig:Hierarchy3} the values of the average clustering coefficients are also given, where the clustering coefficient of a given node is the ratio of the number of existing “triangles” divided by the number of all possible triangles. Here a triangle is made of three edges: two originating from the given node and one connecting its two given neighbors.

In summary, we have carried out a research which was aiming at identifying the typical differences between Horicsanythe three major kinds of forums on the Internet: dark, semi-dark and public (as defined in the Forums section). We considered doing this investigation (being the first of its kind), because obtaining information from several online sources has become one of the major resources of knowledge, and among such texts, the comments of others in the Internet forums play an increasing role. This practice is very important from the point of progress since  the recently introduced variants of social media platforms have greatly contributed to the efficiency of sharing opinions, visual materials resulting in influencing each other as well as in a more efficient communication of ideas and facts. It is widely known that many forums with extremely controversial topics and contents (including those which radicalize the readers or spread information about dangerous products and ideas, e.g., drugs, weapons or aggressive ideologies) exist on the dark web. Therefore, we aimed at comparing the features of this "secretive" version with those of the public web used by the majority of people. We have assumed that a better understanding of the structure, the dynamics and the logic of various kinds of forums is likely to provide us useful information about the laws governing human behavior in this segment of the Internet.

We used network theoretical to analyze the data obtained by studying the connectivity features of the members and the threads within a wide selection of forums (including the dark and the semi-dark ones) and established several characteristic structural features. In addition, we calculated the time dependence of a few quantities of interest associated with the behavior of the commenters. One of the new aspects of our work is that we took into account the directedness of connections among the nodes considered (commenters or threads). We succeeded in identifying a few typical differences between public and dark interactions among the commenters. Our findings reveal both common and rather different features in the two types of behavior. If we locate a behavior that is the same in both the dark and the public web, it may indicate that the type/aspect of the communication we consider is universal, i.e., it is independent from the kind of web used to deliver it. However, we have also demonstrated that several of the various distributions of quantities, like the activity of the commenters, the dynamics of the threads (defined using their lifetime) or the degree distributions corresponding to the three major types of forums, display characteristic deviations. We have also observed a different behavior regarding the number of posts versus the number of their views in the dark and the public forums. We have shown that the networks corresponding to the following two cases (i) the root commenter is regarded as a unique node, or, alternatively, (ii) the root commenter becomes part of the network (so that the related nodes and edges of his/her comments become part of the full network) in many cases result in significantly deviating network structures. Locating characteristic differences can be useful, i.e., when an activity typical for the dark web appears in the public web (as the public web can be accessed and used much more easily), indicating suspicious content. 

\section*{Acknowledgements}
This research was supported by the grant EU FP7 RED-Alert No: 740688.  We acknowledge further partially support by the Bolyai János Research Scholarship, the Bolyai+ Research Scholarship, the ÚNKP-18-1 New National Excellence Program of the Ministry of Human Capacities and the National Research, Development and Innovation Office grant No K128780.
\section*{References}
\bibliographystyle{unsrt}
\bibliography{sample}

\end{document}